\documentclass[11pt]{4ssmmp} 

\def\pa{\partial}

\newcommand{\initiate}{\setcounter{equation}{0}}

\newcommand{\beq}{\begin{equation}}
\newcommand{\eeq}{\end{equation}}
\newcommand\be{\begin{equation} }
\newcommand\bea{\begin{eqnarray}}
\newcommand\ee{\end{equation}}
\newcommand\eea{\end{eqnarray}}

\def\Tr{{\rm Tr}\,}

\usepackage{epsfig}

\newlength{\myVSpace}
\def\endtitle{\par\end{quotation}\vskip3.5in minus2.3in\newpage}

\def\e{\epsilon}

\def\m{\mu}          \def\n{\nu}
       
       \def\r{\rho}
\def\s{\sigma}

         \def\G{\Gamma}

\def\ca{{\cal A}}

\def\cg{{\cal G}}      
      
      \def\cl{{\cal L}}

      \def\car{{\cal R}}

\usepackage{fancyhdr}    
\usepackage{graphicx}
\usepackage{rotating}

\begin{document}
 \baselineskip=11pt

\title{The one-loop renormalization of the gauge sector in
  the $\theta$-expanded noncommutative standard model\hspace{.25mm}\thanks{\,Talk delivered by Voja Radovanovic at '4th Summer School
in Modern Mathematical Physics', Belgrade, Sept. 3-14, 2006,
published in SFIN (year XX) No.A1, 159 (2007); Editors B. Dragovic,
Z. Rakic; Institute of Physics Belgrad, as Proceedings of the 4th
Summer School, Belgrad, Serbia. }}
\author{\bf{Voja Radovanovi\' c}\hspace{.25mm}\thanks{\,e-mail address:
rvoja@phy.bg.ac.yu}
\\ \normalsize{Faculty of Physics, University of Belgrade} \vspace{2mm} \\ \bf{Maja Buri\' c}\hspace{.25mm}\thanks{\,e-mail
address: majab@phy.bg.ac.yu}
\\ \normalsize{Faculty of Physics, University of Belgrade} \vspace{2mm} \\ \bf{Josip Trampeti\' c}\hspace{.25mm}\thanks{\,e-mail
address: josipt@rx.irb.hr}
\\ \normalsize{Institut Rudjer Bo\v skovi\'c, Zagreb, Croatia } }

\date{}

\maketitle

\begin{abstract}
\vspace{2cm} 

In this paper we construct a version of the standard model gauge
sector on noncommutative space-time which is one-loop renormalizable
to first order  in the expansion in the noncommutativity parameter
$\theta$.
\end{abstract}

\clearpage 

 {\Large{ \section{Introduction}

The interest to formulate a consistent quantum field theory on
noncommutative space comes, besides from string theory, also from
mathematics \cite{Kontsevich:1997vb} and from phenomenology. There
are two main approaches  to define
 gauge theories on the canonical noncommutative space. One possibility,
 extensively analyzed in the literature
 \cite{U1,nonab},
 is to replace the ordinary product in the Lagrangian by the Moyal-Weyl
$\star$-product; it is well defined owing to associativity and the
trace property of the $\star$-product. Using this prescription,
however, only U(N) gauge theories can be consistently defined and
the group representations are restricted to
  the  fundamental and the adjoint. This implies  in particular  the
quantization of the electric charge which takes values in $\{ \pm
1 ,0\} $. In perturbative quantization, the interaction vertices
  obtain additional phase factors in comparison with commutative theory,
  and this leads to the well-know UV/IR mixing.

  A slightly different and nonequivalent representation  is
the so-called $\theta$-expanded approach. A consequence of the
requirement that the gauge algebra
 closes on noncomutative fields is
 that the fields are enveloping algebra-valued.
 Using the Seiberg-Witten map,
 which is also an expansion in the noncommutativity parameter $\theta$,
 noncommutative fields are  expressed in terms
of their commutative counterparts \cite{Seiberg:1999vs,Wess}. The
major advantage of this approach is that models with any gauge group
and any particle content can be constructed.

There is a number of versions of the noncommutative standard model
in the $\theta$- expanded approach
\cite{Calmet:2001na,Aschieri:2002mc,Goran,Blazenka}. The argument of
renormalizability was previously not included in the construction
because it was believed that field theories on noncommutative
Minkowski space were not renormalizable in general
\cite{Wulkenhaar:2001sq, Maja}. However, a recent result on the
one-loop renormalizability of the $\theta$-expanded noncommutative
$\rm SU(N)$ gauge theory opens different perspectives
\cite{Buric:2005xe}. Of course, renormalizability in linear order
does not mean renormalizability of the complete theory, but one can
expect that the additional Ward identities, which correspond to the
full noncommutative symmetry and relate different orders, might
help. In this paper we will follow papaer \cite{brt}. We show that
it is possible to construct a version of the NCSM gauge sector which
is one-loop renormalizable to first order in $\theta$.

\initiate \section{ Noncommutative standard model}

\subsection{General considerations}

The noncommutative space which we consider is the flat Minkowski
space, generated by four hermitian coordinates $\widehat x^\mu$
which satisfy the commutation rule
\begin{equation}
[\widehat x^\mu,\widehat x^\nu ] =i\theta^{\mu\nu}={\rm const}.
\label{Mink}
\end{equation}
The algebra of the functions $\widehat\phi (\widehat x)$,
$\widehat \chi (\widehat x)$ on this space can be represented by
the algebra of the functions $\widehat \phi (x)$, $\widehat \chi
(x)$ on the commutative $ {\bf R}^4$ with the Moyal-Weyl
multiplication:
\begin{equation}
\label{moyal} \widehat\phi (x)\star \widehat\chi (x) =
      e^{\frac{i}{2}\,\theta^{\m\n}\frac{\pa}{\pa x^\m}\frac{\pa}{ \pa
      y^\n}}\widehat \phi (x)\widehat \chi (y)|_{y\to x}\ .
\end{equation}
It is possible to represent the action of an arbitrary Lie group
$G$ (with the generators denoted by $T^a$) on noncommutative
space. In analogy to the ordinary  case, one introduces the gauge
parameter $\widehat \Lambda ( x)$ and the vector potential
$\widehat V_\m( x)$.  The main difference is that the
noncommutative $\widehat \Lambda$ and $\widehat V_\m $  cannot
take values in the Lie algebra $\cg$ of the group $G$:  they are
enveloping algebra-valued. The noncommutative gauge field strength
$ \widehat F_{\mu\nu}$ is defined in the usual way
\begin{equation}
\widehat F_{\mu\nu} = \pa_\mu\widehat V_\nu - \pa_\nu \widehat
V_\mu - i(\widehat V_\mu\star\widehat V_\nu - \widehat
V_\nu\star\widehat V_\mu ).                         \label{f}
\end{equation}
There is, however, a relation between the noncommutative gauge
symmetry and the commutative one: it is given by the Seiberg-Witten
(SW) mapping \cite{Seiberg:1999vs}. The expansions of the NC vector
potential and of the field strength, up to  first order in $\theta$,
read \bea && \widehat V_\r(x) =V_\r(x) -\frac 14 \,\theta
^{\m\n}\left\{ V_\m(x), \pa _\n V_\r(x) +F_{\n \r}(x)\right\} .
\nonumber\\
&& \widehat F_{\r\s} =   F_{\r\s} +\frac{1}{4}\theta^{\m\n}
\Big(2\{F_{\m\r},F_{\n\s}\} -\{ V _\m ,(\pa_\n +D_\n )F_{\r\s}
\}\Big)
\nonumber\\
 \eea $D_{\mu}$ is the commutative covariant derivative.

Taking the action of the  noncommutative gauge theory
\begin{equation}
S=-\frac{1}{2}\Tr \int d^4x\,\widehat F_{\m\n}\star\widehat
F^{\m\n} , \label{action}
\end{equation}
and expanding the fields via SW map, $*$ product we obtain the
expression
\begin{equation}
S =-\frac 12\Tr\int d^4x\,F_{\m\n}F^{\m\n}+\theta^{\m\n}\,\Tr\int
d^4x\, \big(\frac 14 \, F_{\m\n}F_{\r\s}- F_{\m\r}F_{\n\s}
\big)F^{\r\s} . \label{Act}
\end{equation}

\subsection{$\rm U(1)_Y\otimes SU(2)_L\otimes SU(3)_C$}

The discussion given above was a general one, without any
specification of the gauge group $G$ or of its representations. In
(\ref{Act}) we    have a factor $\Tr \{T^a,T^b\}T^c \sim d^{abc}$.
One could perhaps assume that, as the field strength transforms
according to the adjoint representation, the symmetric
coefficients $d^{abc}$ are given in that representation. However,
when the matter fields are included, other representations of $G$
are present too, and therefore the expression (\ref{Act}) is
ambiguous.

To start the discussion of the gauge field action-dependence on the
gauge group and/or on its representation, we use the most general
form of the action, \cite{Aschieri:2002mc}: \be S_{cl} =
-\frac{1}{2}  \int d^4x \, \sum_{\car} {C_{\car}} {\Tr}\Big(
{\car}(\widehat F_{\mu \nu}) * {\car}(\widehat F^{\mu \nu})\Big).
\label{action1} \ee The sum is, in principle, taken  over all
irreducible representations $\car$  of $G$  with arbitrary weights
$C_\car$. Of course, for the gauge group $G$ we take $\rm
U(1)_Y\otimes SU(2)_L\otimes SU(3)_C$. The previous action may be
generalized by adding $x^2-$depending term \bea S_{cl}& =&
-\frac{1}{2} \int d^4x \, \sum_{\car} {C_{\car}} {\Tr}\Big(
{\car}(\widehat F_{\mu \nu}) * {\car}(\widehat F^{\mu
\nu})\Big)\nonumber\\&+&\frac{a-1}{4}\int
d^4x\,\sum_{\car}{\Tr}\Big(h\theta^{\r\s}\star\widehat
{\car}(\widehat F_{\r \s})\star {\car}(\widehat (F_{\mu \nu})\star
{\car}(\widehat F^{\mu \nu})\Big). \nonumber \eea The above action
has very interesting renormalization property and phenomenological
consequences. The constant $a$ will be fixed by renormalizability
property. For $a=1$ we obtain (\ref{action1}), so-called minimal
model.

  To relate
the action (\ref{Act}) to the usual action of the commutative
standard model, we make the decompositions \bea V_\m &=& g^\prime
\ca _\m \car (Y) + g B_\m^i \car (T^i_L) +g_S G_{\m}^a\car
(T^a_S),\nonumber\\
F_{\mu\nu} &=&g^\prime f_{\mu\nu}\car(Y)+g B_{\m\n}^i\car
(T^i_L)+g_S G_{\m\n}^a\car(T^a_S).\nonumber\eea The $\car(Y)$,
$\car (T_L^i)$,  $\car (T_S^a)$ denote the representations of the
group generators $Y$, $T_L^i$ and $T_S^a$ of $\rm U(1)_Y$, $\rm
SU(2)_L$ and $\rm SU(3)_C$, respectively; the group indices run as
$i,j = 1,\dots 3$ and  $a,b = 1,\dots 8$. According to
\cite{Aschieri:2002mc}, we take that  $C_\car $ are nonzero only
for the particle representations which are present in the standard
model. Then from (\ref{action1}) we obtain the expression for the
$\theta$-independent part of the Lagrangian
\begin{eqnarray}
\cl_{SM} &=&-\frac{g^{\prime 2}}{2}\sum_{\car}C_{\car}
d(\car_2)d(\car_3)\car_1(Y)\car_1(Y)\,f_{\m\n}
f^{\m\n}\nonumber \\
&-&\frac{g^2}{2}\sum_{\car}C_{\car}d(\car_3)\Tr(\car(T_L^i)\car(T_L^j))\,
B_{\m\n}^i
B^{\m\n j}\nonumber\\
&-&\frac{g^2_S}{2}\sum_{\car}C_{\car}d(\car_2)\Tr(\car(T_S^a)\car(T_S^b))\,
G_{\m\n}^a G^{\m\n b}, \label{LSM}
\end{eqnarray}
where $d(\car)$ denotes the dimension of the representation
$\car$. The noncommutative correction, that is the $\theta$-linear
part of the Lagrangian, reads\bea \cl^\theta &=& \sum \cl^\theta
_i =g^{\prime 3}\kappa_1\theta^{\m\n} \left( \frac a4
f_{\m\n}f_{\r\s}f^{\r\s}-f_{\m\r}f_{\n\s}f^{\r\s}\right)
\nonumber\\
&+&g^3 \kappa_4^{ijk}\theta^{\m\n}
 \left(\frac a4 B_{\m\n}^i B_{\r\s }^j B^{\r\s k} - B_{\m\r}^i B_{\n\s }^j
B^{\r\s k} \right) \nonumber\\
&+& g^3_S\kappa_5^{abc}\theta^{\m\n}
 \left( \frac{ a}{4}G_{\m\n}^a G_{\r\s }^b G^{\r\s c}- G_{\m\r}^a G_{\n\s }^b
G^{\r\s c}\right) \nonumber \\
&+& g^\prime g^2  \kappa_2 \theta^{\m\n} \left(\frac a4
f_{\m\n}B_{\r\s }^i B^{\r\s i} -f_{\m\r}B_{\n\s }^i B^{\r\s i} +
c.p.\right) \nonumber \\
&+& g^\prime g^2_S  \kappa_3 \theta^{\m\n} \left(\frac a4
f_{\m\n}G_{\r\s }^a G^{\r\s a}\right.\nonumber\\&
-&\left.f_{\m\r}G_{\n\s }^a G^{\r\s a} +c.p. \right),\label{lterm}
\eea
where the $c.p.$ in (\ref{lterm}) denotes the addition of the
terms obtained by a cyclic permutation of fields without changing
the positions of indices. The couplings in (\ref{lterm}) are
defined as follows: \bea \kappa_1 &=&\sum_{\car}C_{\car}
d(\car_2)d(\car_3)\car_1(Y)^3,
\nonumber \\
\kappa_2 \delta^{ij} &=& \sum_{\car}C_{\car} d(\car_3)\car_1
(Y)\Tr (\car_2(T^i_L)\car_2(T^j_L)),
\nonumber \\
\kappa_3 \delta^{ab}&=& \sum_{\car}C_{\car} d(\car_2)\car_1 (Y)\Tr
(\car_3(T^a_S)\car_3(T^b_S)),
\nonumber \\
\kappa_4 ^{ijk}&=& \frac12\sum_{\car}C_{\car} d(\car_3)\Tr
(\{\car_2(T^i_L),\car_2(T^j_L)\}\car_2(T^k_L)),
\nonumber \\
\kappa_5^{abc} &=&\frac12 \sum_{\car}C_{\car} d(\car_2)\Tr
(\{\car_3(T^a_S),\car_3(T^b_S)\}\car_3(T^c_S)) . \nonumber \eea

Let us  discuss the dependence of $\kappa_1, \dots  ,\kappa_5$ on
the representations of matter fields. For the first generation of
the standard model there are six such representations ; they
produce six independent constants $C_\car$\footnote{We assume that
$C_\car >0$; therefore the six $C_\car$'s were denoted by
 $\frac{1}{g_i^2}\,,i=1,...,6$, in \cite{Calmet:2001na,Goran}.}.
These constants  are constrained by the three relations which
defined $g^\prime,g,g_S$. One can immediately verify that
$\kappa_4^{ijk} = 0$.  We shall in addition take that
$\kappa^{abc}_5 = 0$. The argument for this assumption is related to
the invariance of the colour sector of the SM  under charge
conjugation. Although apparently  one has only the fundamental
representation {\bf 3} of $\rm SU(3)_C$, there are in fact both
${\bf 3}$ and ${\bf \bar 3}$ representations with the same weights,
$C_{{\bf 3}} =C_{{\bf\bar 3}}$. Since the symmetric coefficients for
the {\bf 3} and ${\bf \bar 3}$ representations satisfy
$d^{abc}_{{\bf \bar 3}} = - d^{abc}_{{\bf 3}}$, we obtain
\begin{equation}
 \kappa_5^{abc} = C_{{\bf 3}}d_{\bf 3}^{abc} + C_{{\bf \bar 3}}d_{\bf \bar
3}^{abc} =0 .\label{Kappa}
\end{equation}

We are left only with three non vanishing couplings, $\kappa_1$,
$\kappa_2$ and $\kappa_3$, depending on six  constants $C_1,\dots
,C_6.$ Our classical noncommutative action reads \cite{brt} \be
S_{cl} = S_{SM} +S^\theta , \label{Scl} \ee with \bea S^\theta &=&
g^{\prime 3}\kappa_1\theta^{\m\n} \int d^4 x\left( \frac a4
f_{\m\n}f_{\r\s}f^{\r\s}-f_{\m\r}f_{\n\s}f^{\r\s}\right)\nonumber\\
&+& g^\prime g^2  \kappa_2 \theta^{\m\n}\int d^4 x\left(\frac a4
f_{\m\n}B_{\r\s }^i B^{\r\s
i} -f_{\m\r}B_{\n\s }^i B^{\r\s i} + c.p.\right) \nonumber \\[4pt]
&+& g^\prime g^2_S \kappa_3 \theta^{\m\n}\int d^4 x\left(\frac a4
f_{\m\n}G_{\r\s }^a G^{\r\s a}\right.\nonumber\\&
-&\left.f_{\m\r}G_{\n\s }^a G^{\r\s a} +c.p. \right). \label{St}
\eea
\smallskip

The first term  in (\ref{St})  is one-loop renormalizable to linear
order in $\theta$ \cite{Buric:2005xe} since the one-loop correction
 is of the second order in $\theta$. We need to investigate only
the renormalizability of remaining  parts of the action (\ref{St}).

\section{One-loop renormalizability}

We compute the divergences in the one-loop effective action using
the background-field method.  Here we will give only main results,
the details are given in \cite{brt}. For the action (\ref{Act}), the
classical Lagrangian reads \bea \cl_{cl} &=& -\frac 14
f_{\m\n}f^{\m\n} - \frac 14 B_{\m\n}^i B^{\m\n i} -\frac
14G_{\m\n}^a G^{\m\n a}
 \label{lagr}\\
&+& g^{\prime 3}  \kappa_1 \theta^{\m\n} \left(\frac a4
f_{\m\n}f_{\r\s } f^{\r\s } -f_{\m\r}f_{\n\s } f^{\r\s } \right)
\nonumber\\
&+& g^\prime g^2  \kappa_2 \theta^{\m\n} \big(\frac a4
f_{\m\n}B_{\r\s }^i B^{\r\s i} - 2f_{\m\r}B_{\n\s }^i B^{\r\s
i}\nonumber\\& +& \frac a2 f_{\r\s}B_{\m\n }^i B^{\r\s i} -
f_{\r\s}B_{\m\r }^i B^{\n\s i}
 \big)
\nonumber\\
&+& g^\prime g^2_S  \kappa_3 \theta^{\m\n} \big(\frac a4
f_{\m\n}G_{\r\s }^a G^{\r\s a} - 2f_{\m\r}G_{\n\s }^a G^{\r\s
a}\nonumber\\& +& \frac a2 f_{\r\s}G_{\m\n }^a G^{\r\s a} -
f_{\r\s}G_{\m\r }^a G^{\n\s a}
 \big) \,,
 \nonumber
\eea
 After a very long and straightforward calculation \cite{brt}
we get the diveregent part of one-loop effective action
 \bea \G_{div}&=& \frac{1}{3(4\pi)^2\epsilon}11\Big[\int d^4 x
B_{\m\n}^iB^{\m\n i} + \frac {33}{2}\int d^4 x G_{\m\n}^aG^{\m\n
a}
\nonumber \\
&+& 4(3-a)g^\prime g^2 \kappa_2\theta^{\m\n}\int d^4 x \big(\frac
14 f_{\m\n}B_{\r\s }^i B^{\r\s i} - f_{\m\r}B_{\n\s }^i B^{\r\s i}
 \big)
 \nonumber \\
&+& 6(3-a)g^\prime g^2_S \kappa_3\theta^{\m\n}\int d^4 x\big(\frac
14 f_{\m\n}G_{\r\s }^a G^{\r\s a} \nonumber\\&-& f_{\m\r}G_{\n\s
}^a G^{\r\s a}
 \big)\Big].
 \label{div}
 \eea
The divergent contribution due to $\rm U(1)_Y$ solely vanishes, both
the commutative and the noncommutative one. It is clear from
(\ref{div})  that the divergences in the noncommutative sector
vanish for the choice $a=3$. Therefore one obtains that the
noncommutative gauge sector interaction is not only renormalizable
but finite. The renormalization is performed by adding  counter
terms to the Lagrangian. We obtain
 \bea \cl +\cl_{ct}&=&
-\frac{1}{4}{f_0}_{ \m\n }{f_0}^{\m\n }-\frac{1}{4}{B_0}_{ \m\n
}^i{B_0}^{\m\n i} -\frac{1}{4}{G_0}_{ \m\n }^a{G_0}^{\m\n a}
\nonumber\\
&+& g^{\prime 3}\kappa_1\theta^{\m\n} \left( \frac 34
{f_0}_{\m\n}{f_0}_{\r\s}{f_0}^{\r\s}-{f_0}_{\m\r}{f_0}_{\n\s}{f_0}^{\r\s}\right)
\nonumber\\
&+& g_0^\prime g_0^2  \kappa_2 \theta^{\m\n} \left(\frac 34
{f_0}_{\m\n}{B_0}_{\r\s }^i B_0^{\r\s i} -{f_0}_{\m\r}{B_0}_{\n\s
}^i B_0^{\r\s i}+c.p.\right)
 \nonumber \\
&+& g_0^\prime (g_{S})_0^2  \kappa_3 \theta^{\m\n} \left(\frac 34
{f_0}_{\m\n}{G_0}_{\r\s }^a G_0^{\r\s
a}\right.\nonumber\\&-&\left.{f_0}_{\m\r}{G_0}_{\n\s }^a G_0^{\r\s
a}+c.p.\right), \label{lct} \eea where the bare quantities are given
as follows: \bea {\ca_0}^{\m}&=&\ca^{\m }\, ,\qquad g_0 ^\prime = g
^\prime\, ,
\label{A0}\\
{B_0}^{\m i}&=&B^{\m i}\sqrt{1+\frac{44g^2}{3(4\pi)^2\epsilon}}\,
,\quad g_0
=\frac{g\,\m^{\e/2}}{\sqrt{1+\frac{44g^2}{3(4\pi)^2\epsilon}}}\, ,
\label{B0}\\
{G_0}^{\m a}&=&G^{\m
a}\sqrt{1+\frac{22g_S^2}{(4\pi)^2\epsilon}}\,,\quad{(g_{S})}_0=\frac{g_S\,\m^{\e
/2}}{\sqrt{1+\frac{22g_S^2}{(4\pi)^2\epsilon}}}\,  .\label{G0}
 \eea

Finally, an important point is that the noncommutativity parameter
$\theta$ need not be renormalized.

\section{Discussion and conclusion}

We have constructed  a version of the standard model on the
noncommutative Minkowski space which is one-loop renormalizable and
finite in the gauge sector and in first order in the $\theta$
parameter. The renormalizability in the model was obtained by
choosing  six particle representations of the matter fields for the
first generation of the SM, and by fixing the parameter $a=3$.

The one-loop renormalizability of the NCSM gauge sector is certainly
a very encouraging result from both  theoretical and experimental
perspectives. So far fermions have not been successfully included:
the results on the renormalizability of noncommutative gauge
theories with Dirac fermions are negative
\cite{Wulkenhaar:2001sq,Maja} as a $4\psi$-divergence always
appears. In the case of SU(N) or SU(3)$\otimes$SU(2)$\otimes$U(1)
the unexpanded gauge theory cannot be consistently defined.
Furthermore, our results show that the requirement of
renormalizability fixes the parameter $a$ to $a=1$ or $a=3$
\cite{new}. We hope that a similar procedure could be applicable to
the fermionic sector of the theory.

}}

%
%
%
%
%
%

\end{document}